\documentclass[twocolumn,aps,prl,10pt,superscriptaddress]{revtex4-1} % For final two-column format

\usepackage{graphicx}% Include figure files
\usepackage{amsmath,amssymb}
\usepackage{esvect}
\usepackage{booktabs}

\renewcommand{\vec}[1]{\ensuremath{\mathbf{#1}}}

\begin{document}

\title{Search for Screened Interactions Associated with Dark Energy \\ Below the 100~$\mathrm{\mu m}$ Length Scale}
   
\author{Alexander D. Rider}
\affiliation{Department of Physics, Stanford University, Stanford, California 94305, USA}
\author{David C. Moore}
\affiliation{Department of Physics, Stanford University, Stanford, California 94305, USA}
\author{Charles P. Blakemore}
\affiliation{Department of Physics, Stanford University, Stanford, California 94305, USA}
\author{Maxime Louis}
\affiliation{Department of Physics, \'{E}cole Polytechnique, 91128 Palaiseau, France}
\author{Marie Lu} 
\affiliation{Department of Physics, Stanford University, Stanford, California 94305, USA}
\author{Giorgio Gratta}
\affiliation{Department of Physics, Stanford University, Stanford, California 94305, USA}

\date{\today}
\begin{abstract}
We present the results of a search for unknown
interactions that couple to mass between an optically levitated microsphere and a gold-coated silicon cantilever. The scale and geometry of the apparatus enables a search for new forces that appear at distances below 100~$\mu$m and which would have evaded previous searches due to screening mechanisms. The data are consistent with electrostatic backgrounds and place upper limits on the strength of new interactions at $<0.1$~fN in the geometry tested. For the specific example of a chameleon interaction with an inverse power law potential, these results exclude matter couplings $\beta > 5.6 \times 10^4$ in the region of parameter space where the self-coupling $\Lambda \gtrsim 5$~meV and the microspheres are not fully screened.

\end{abstract}

\maketitle

Observations indicate that the universe is expanding at an accelerating rate~\cite{Riess98,Perlmutter98,Ade15}. 
This acceleration can be explained by the presence of `dark energy' throughout the universe~\cite{Agashe2014}.  Although the nature of dark energy is unknown, one possibility is that it consists of a scalar field that couples to mass~\cite{Jain10,Joyce14}. Astrophysical measurements of the dark energy density imply an energy scale of $\Lambda = 2.4$~meV, corresponding to a length scale of $\hbar c$/$\Lambda \sim80 \, \mu$m.

It might be possible to detect the presence of a scalar field constituting dark energy by searching for new interactions between objects separated by distances below the dark energy length scale~\cite{Beane97,adelberger07,Jain10,Joyce14}.  In many cases, the resulting forces can be substantially larger than Newtonian gravity at short distances~\cite{Joyce14,Mota06}. The most sensitive previous searches for violations of Newtonian gravity at or below the dark energy length scale employed macroscopic test masses or a conductive shield between the probe and test masses to minimize electromagnetic backgrounds~\cite{Adelberger09,adelberger07,Geraci08,Sushkov11,Wen16}. 
Although these experiments place stringent constraints on deviations from Newtonian gravity, it is possible to construct theories of dark energy involving new forces that could have avoided detection due to the geometry and scale of previous experiments~\cite{Mota06,Upadhye12,Burrage15,Joyce14}.  For these screened interactions, recent searches using microscopic test masses such as atoms~\cite{muller15,Elder16} or neutrons~\cite{Li16,Lemmel15,Jenke14} often provide the strongest constraints.

Several screening mechanisms have been proposed to evade existing experimental constraints on scalar interactions in the laboratory and solar system~\cite{Joyce14}. A specific example is the chameleon mechanism~\cite{Khoury03a,Khoury03b}, in which the effective mass of the chameleon particle (corresponding to the inverse length scale of the interaction) depends on the local matter density. At cosmological distances where the matter density is low, the chameleon field would mediate a long range interaction that explains the accelerating expansion of the universe~\cite{brax04}. However, most laboratory experiments are carried out in regions of high matter density, where the forces arising from the chameleon interaction are suppressed. 

\begin{figure}[t]
\includegraphics[width=\columnwidth]{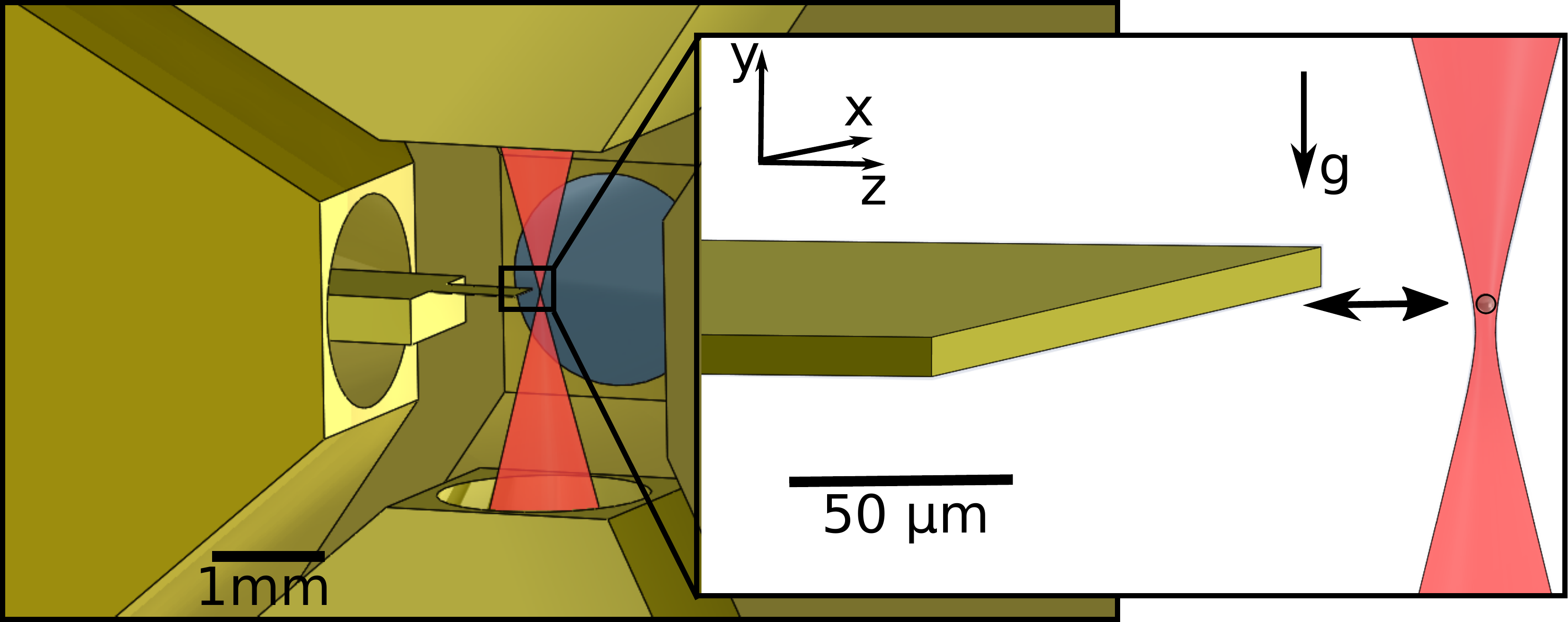}
\caption{(left) Schematic of the optical trap and shielding electrodes. The electrode in the foreground is removed to show the inside of the trap. (right) Zoom in on the region near the trap. A $5\ \mu$m diameter microsphere is suspended at the
focus of an upward propagating laser beam. The $10\ \mu$m thick Au-coated Si cantilever is positioned at $\sim$20\textendash200~$\mu$m separations from the microsphere and oscillated in the $z$ direction using a nanopositioning stage.}
\label{fig:schematic}
\end{figure} 

This work presents a search for screened interactions below the dark energy length scale using optically levitated $\mu$m-size dielectric spheres as test masses.  Levitated microspheres in high vacuum~\cite{Chang10,Romero10,Yin13, Li11,Gieseler12,Kiesel13,Ranjit15,Millen15,Fonseca15} can be used to detect forces $\ll 10^{-18}$~N~\cite{Geraci10, Moore14, Gieseler13, Ranjit15, Ranjit16}, and in many cases their small size avoids screening effects.  

The test masses used in this work consist of amorphous silica microspheres with radius $r = 2.5\ \mu$m and mass $m = 0.13$~ng~\footnote{Bangs Laboratories, Inc., http://www.bangslabs.com} levitated in a single-beam, upward-propagating 1064~nm laser trap~\cite{Ashkin71,Moore14}. The radiation pressure from the laser counters Earth's gravity and acts as an optical spring pulling the microsphere to the center of the Gaussian beam~\cite{ashkin}. The resonant frequencies of the trap are $\sim 150$~Hz for the 2 degrees-of-freedom orthogonal to the Earth's gravity and $\sim 100$~Hz for the degree-of-freedom parallel to Earth's gravity. The position of the microsphere is measured by focusing secondary 650~nm Gaussian laser beams on the microsphere and imaging the pattern of scattered light onto a position-sensitive photodiode (PSPD). For small displacements from the center of the trap, the PSPD produces a differential current that is a linear function of the position of the microsphere.   

When the microspheres are loaded into the optical trap, they typically have an electric charge of $\sim100 e$~\cite{Moore14}. The charge is measured by monitoring the response to an oscillating electric field. Microspheres are discharged with UV radiation from a Xenon flash-lamp. As shown in~\cite{Moore14}, clear charge quantization can be observed at the end of the discharging cycle, providing a force calibration for the system. 

The microspheres are levitated inside of a vacuum chamber to reduce the force noise coming from collisions with residual gas. Due to reduced gas damping, the trap becomes unstable below 0.05~mbar. To stabilize the trap, active feedback is applied by measuring the microsphere's position and modulating the position of the trap. Measurements are performed at pressures below $10^{-6}$~mbar where the noise for force measurement is limited to $2\times10^{-17}\ \mathrm{N\ Hz}^{-1/2}$ by imaging noise. The optical setup and calibration methods are improved versions of those discussed in \cite{Moore14}.

A schematic view of the apparatus is shown in Fig.~\ref{fig:schematic} where a coordinate system is defined.  The microsphere coupling is probed with a silicon cantilever with dimensions $500\ \mu \mathrm{m} \times 2000\ \mu \mathrm{m} \times 10\ \mu \mathrm{m}$ and a 500 $\mu$m thick handle, fabricated from a silicon on insulator (SOI) wafer using optical photolithography and plasma etching. The $10\ \mu m$ dimension is oriented so that the cantilever clears the Gaussian beam waist of the laser and the 500~$\mu$m dimension is approximately centered on the trap in the \textit{x} direction. A 200~nm gold shielding layer was evaporated onto the cantilever to minimize its electrostatic interactions with the microsphere. The cantilever is mounted on a 3-axis nanopositioning stage used to control its spacing from the microsphere with a precision of 3~nm and a travel of $80\ \mu$m~\footnote{Newport, product number: NPXYZ100SGV6, http://www.newport.com}.  The trap and cantilever are electrically shielded inside a cube consisting of six gold-plated electrodes separated by 4~mm, whose potentials are controlled by external digital-to-analog converters (DACs).  The nanopositioning stage is mounted on a piezo motor driven stage with 12~mm travel in the $z$ direction to provide coarse positioning. 

To measure electrostatic interactions between the cantilever and the microsphere, each shielding electrode was set to a nominal potential of 0~V while the cantilever was biased to a non-zero potential. The $z$ position of the nanopositioning stage was driven with an $18.3$~Hz sine wave over its full 80 $\mu$m travel. The microspheres were aligned with the center of the cantilever in the $y$ direction by determining the position at which the maximum electrostatic response was seen as the cantilever was swept in the $z$ direction at fixed bias. The microsphere and stage positions were recorded in 50~s long integrations. Data were acquired for coarse stage positions with closest approach of 20, 60, 100, and 150~$\mu$m.  This procedure was repeated for each of three microspheres considered in this work. 

\begin{figure}[t]  
  \includegraphics[width=\columnwidth]{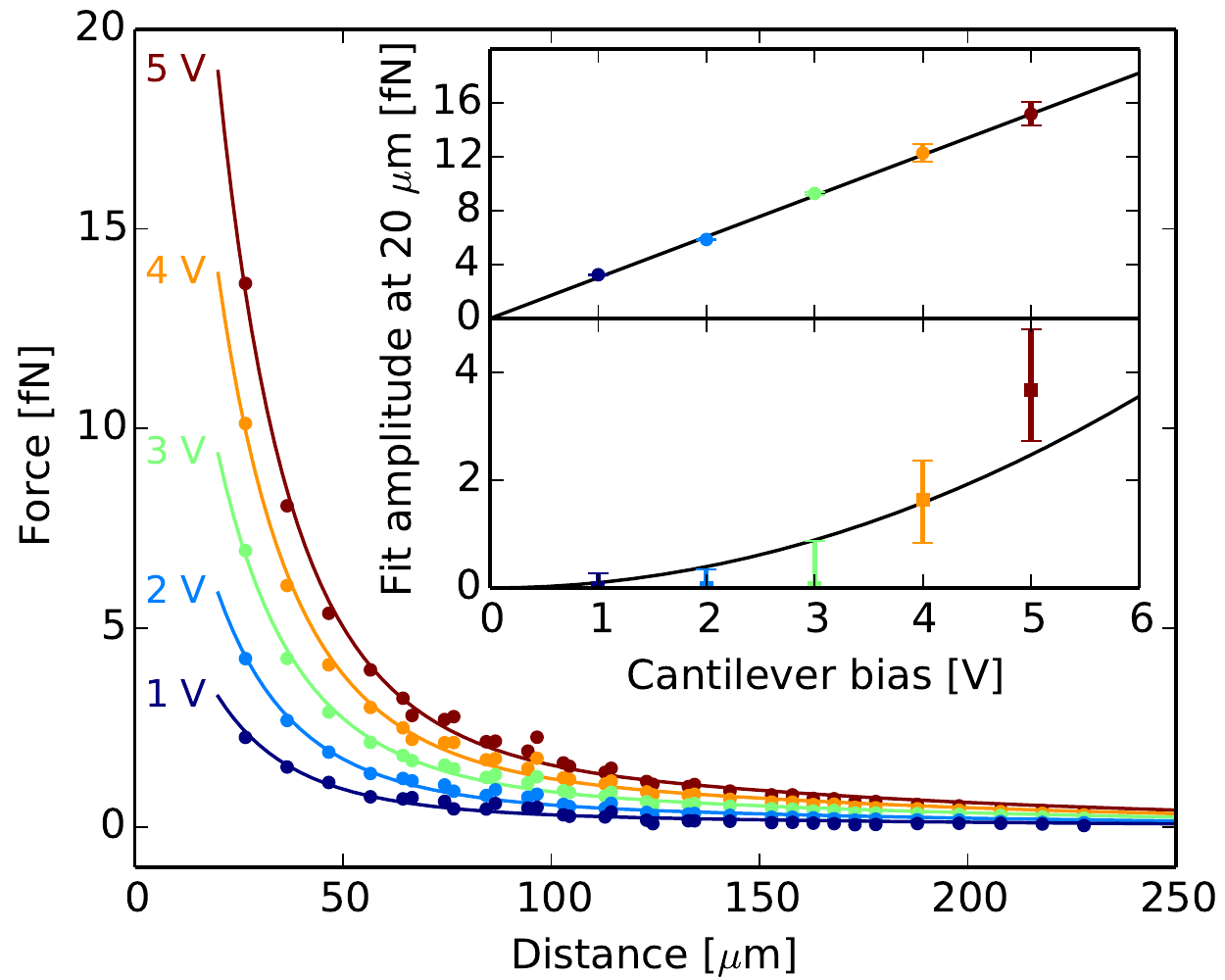}
  \caption{ Measured response of microsphere \#1 versus distance from the
    cantilever face as the cantilever is swept in $z$ with a constant bias of
    1,2,3,4, and 5~V.  The data points are shown as dots and the best fit
    model as solid lines. (inset) Amplitude of the fit
    component $\propto \partial_{z} E_z$ (top) and the fit component
    $\propto E_z \partial_{z} E_z$ (bottom).  Fits to the expected
    linear and quadratic dependence on the voltage are also shown (solid lines).}
\label{fig:calib}
\end{figure}

To eliminate low frequency drifts, the microsphere positions were first mean subtracted. The data were then averaged in 10~$\mu$m cantilever position bins and calibrated to force units using the single-charge-step calibration discussed previously. Data at neighboring coarse-stage positions were matched in the 30\textendash40~$\mu$m overlap region. The measured electrostatic force versus spacing between the cantilever and the microsphere is shown in Fig.~\ref{fig:calib}.

Although electrically neutral microspheres are used, they still contain $\sim10^{14}$ charges and interact primarily as electric dipoles.  The force on a microsphere with dipole moment $\vec{p}$ is given by $\vec{F} = (\vec{p} \cdot \nabla) \vec{E}$~\cite{jackson99}, where $\vec{p} = \vec{p}_0 + \alpha \vec{E}$ consists of a permanent dipole, $\vec{p}_0$, and an induced dipole, $\alpha \vec{E}$, for polarizability $\alpha$. The latter term incorporates any dipole induced by an electric field, including the linear dielectric response as well as any non-zero surface charge mobility. Figure~\ref{fig:calib} shows a fit to the model $\vec{F} \cdot \hat{z} \equiv F_b(z) = (p_{x}\partial_{x} + p_{y}\partial_{y} + p_{z}\partial_{z})E_{z} \approx p_{0z}\partial_{z} E_{z} + \alpha E_{z}\partial_z E_{z}$. 

\begin{table}
  \caption{Dipole moments and polarizabilities measured for each microsphere.}
  \label{tab:dipole_moments}
  \begin{tabular}{ccc}
    \toprule
    Microsphere \, & \, $p_{0z}$ [$e$\,$\mu$m] \, & \, $\alpha/\alpha_0$\\
    \hline
    \hline
    \midrule
    \#1 & 151 $\pm$ 6 & 0.21 $\pm$ 0.13\\
    \#2 & 89 $\pm$ 10 & 0.00 $\pm$ 0.33\\
    \#3 & 192 $\pm$ 30 & 0.25 $\pm$ 0.14\\
    \hline
    \hline
    \bottomrule
  \end{tabular}
\end{table}

A finite-element method (FEM) was used to solve for $\vec{E}$ within the geometry of the trapping region. Dipole moments and polarizabilities were extracted by fitting the microsphere responses at non-zero cantilever bias to $F_b(z)$. The results of this fit for each microsphere are shown in Table~\ref{tab:dipole_moments}.  The dipole moments are measured in units of $e$\,$\mu$m and the polarizabilities are  reported relative to $\alpha_{0} = 3\epsilon_{0}\left(\frac{\epsilon_{r} - 1}{\epsilon_{r} + 2}\right)(\frac{4}{3}\pi r^{3})$ assuming $\epsilon_r \sim3$ and $r = 2.5\ \mu$m. The reported values of polarizability, which are smaller than $\alpha_0$, could arise from systematics in the determination of a small induced dipole on top of a much larger permanent dipole, an unexpectedly low value of $\epsilon_r$, or a smaller than expected volume.

Following the measurement of the electrostatic interaction at a given coarse stage position, the cantilever was set to a nominal potential of 0~V, and twenty additional 50~s long integrations were acquired to search for new screened interactions. This procedure was then repeated to obtain three 1000~s measurements at each coarse stage position in order to quantify time dependent variation in the measured response over a period of several hours. The standard deviation of the repeated measurements at each position bin was included as an additional systematic error.

\begin{figure}[t] 
  \includegraphics[width=\columnwidth]{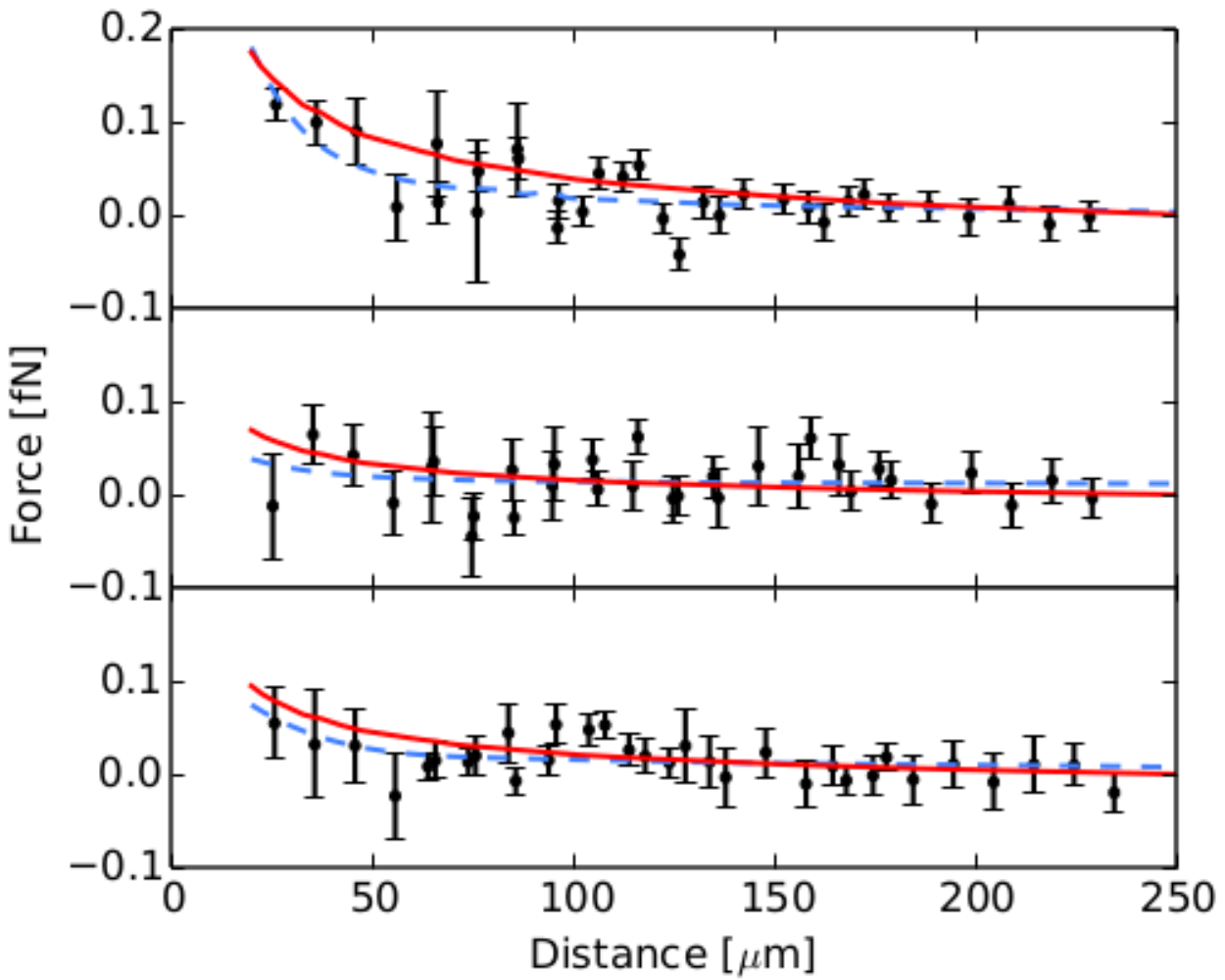}
  \caption{Measured response for microspheres \#1 (top), \#2 (middle), and
    \#3 (bottom) versus distance from the cantilever face as the
    cantilever is swept in $z$ with a nominal bias of 0~V. The
    best fit electrostatic background-only model (dashed) and the amplitude
    of a chameleon force at the 95\% CL upper limit for $\Lambda =
    10$~meV (solid) are also shown.  }
\label{fig:Force_vs_pos}
\end{figure}

The measured force versus position for each of the three microspheres is shown in Fig.~\ref{fig:Force_vs_pos}.  A small residual force $\lesssim 10^{-16}$~N can be seen for each microsphere.  This response is consistent with electrostatic forces resulting from the permanent electric dipole moment of the microspheres coupling to the electric field produced by potential differences between the cantilever and shielding electrodes of $\lesssim 30$~mV.  Contact potentials of this scale are expected to arise between connections to the electrodes in the vacuum chamber and external electronics.

The data shown in Fig.~\ref{fig:Force_vs_pos} can be used to set constraints on new screened interactions at distances of the order of the dark energy length scale, with a sensitivity that is limited by the presence of the residual electrostatic backgrounds.  As a concrete example, we consider the presence of a non-relativistic, steady-state chameleon field, $\phi$, that mediates a force between the microsphere and cantilever. Following \cite{Burrage15,Mota06,Elder16}, we assume an inverse power law form of the effective potential $V_{\mathrm{eff}}(\phi) = \Lambda^{4}[1+(\Lambda / \phi)^n] + (\beta \rho / M_{Pl} ) \phi $. Here, $\Lambda$ is the scale of the chameleon self interaction, often chosen at the dark energy scale $\Lambda \sim2.4$~meV. The coupling to matter of density $\rho$ is determined by the scale $M = M_{Pl}/\beta$ where $M_{Pl}$ is the reduced Planck mass and $\beta$ is unitless. Although other power laws are possible, $n=1$ was chosen as a characteristic example for this search.

Similar to the electric field calculation described above, an FEM was employed to solve the non-linear equation of motion $ \mathbf{\nabla}^2 \phi = \partial_{\phi} V_{\mathrm{eff}}$ in the geometry described previously. The residual gas pressure of $\sim10^{-6}$ mbar was included, but has negligible effect on $\phi$ for values of the matter coupling considered in this work. Boundary conditions were set to the equilibrium value of the field within the cantilever and electrodes, following the detailed treatment of matter-vacuum interfaces in \cite{Elder16}.

The resulting chameleon force on a microsphere in the $z$ direction was calculated as $F_c(z, \beta, \lambda) = \lambda (\beta \rho / M_{Pl}) \int_{V}  (\partial_{z} \phi) \, dV $ where $\rho$ and $V$ are the density and volume of the microsphere and $\lambda$ is a screening factor~\cite{Burrage15,Elder16}.  In the region of parameter space where $\rho r^2 < 3 M_{Pl}\phi/\beta$, the microsphere is unscreened and $\lambda = 1$. However, when $\beta$ becomes sufficiently large, the force on the microsphere is suppressed by $\lambda < 1$~\cite{Burrage15,Elder16}. 

The data for each microsphere were fit to a model $F(z) = A_{c}F_{c}(z,\beta,\Lambda) + A_{b}F_{b}(z) + A_0$, where $F_{b}(z)$ is the shape of the empirical background measured for each microsphere when the cantilever was biased to $5$~V, $A_{b}$ is the unknown electrostatic background amplitude due to residual contact potentials on the electrodes, and $A_0$ accounts for the arbitrary offset subtracted from the data at each coarse stage position. 

$A_{c}$, the normalization of the chameleon force, was constrained in the fit by the following systematics. The microsphere mass was not directly measured, but the radius of the spheres was determined by the manufacturer to be $2.5 \pm 0.24\ \mu$m, leading to a 35\% uncertainty on the chameleon force.  Fits of the calibration data to the electric field simulations indicate that the microsphere was centered in $y$ relative to the cantilever within $4\ \mu$m, leading to an uncertainty on the amplitude of the chameleon force of $1.8$\%.  The $z$ position of the coarse stage was determined from microscope images of the cantilever to $\lesssim10\ \mu$m, at each coarse stage setting. Using the positions and uncertainties determined from the calibration images, the coarse stage positions were further refined by allowing $z$-position offsets to float at each coarse stage position in the electrostatic fit. The best fit positions were used in the final chameleon fit, and their uncertainties contribute an additional systematic error of 6\%.  All errors were added in quadrature for a total systematic error of 36\% on $A_{c}$, dominated by the uncertainty in the microsphere masses.

At each value of $\Lambda$, the profile of the negative log likelihood (NLL) was calculated by minimizing the NLL for the fit at each value of $\beta$ over the nuisance parameters $A_{c}$ (including its Gaussian constraint), $A_{b}$ and $A_0$.  The 95\% confidence interval for $\beta$ was determined from the combined profile from all three microspheres following Wilks' theorem~\cite{wilks38,Cowan98}. This was done assuming that 2NLL follows a $\chi^{2}$ distribution with one degree-of-freedom (DOF). The $\chi^{2}$ statistic at the best fit point and for the background only model indicates that both provide a good fit to the data. At the best fit point, $\chi^{2}=97.8$ for 87 DOF, while for the background only model $\chi^{2} = 98.9$ for 88 DOF. For all $\Lambda$, the data are consistent with the background-only model at the 95\% confidence level (CL).  The background-only fits are shown in Fig.~\ref{fig:Force_vs_pos}, together with the amplitude of a chameleon force at the 95\% CL upper limit. 

\begin{figure}[t] 
  \includegraphics[width=\columnwidth]{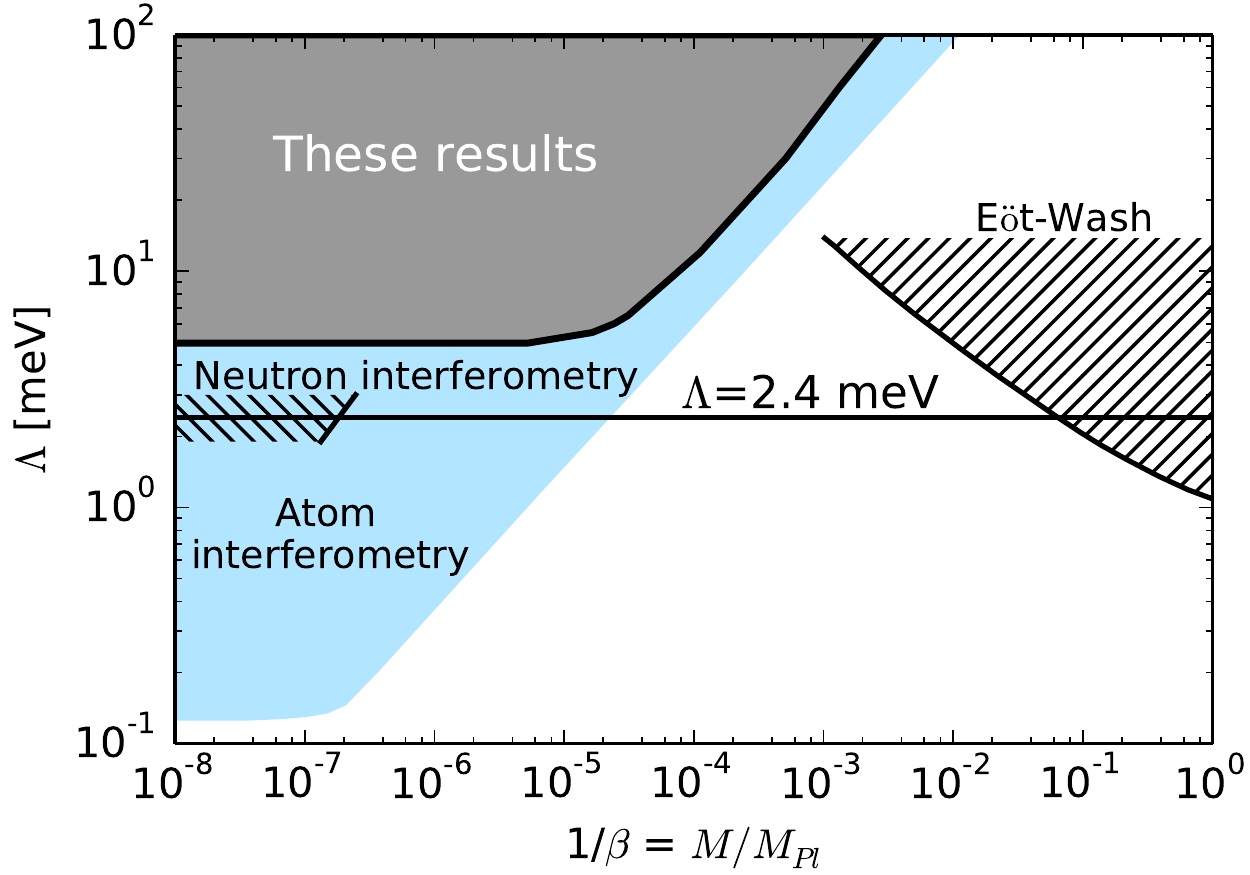}
  \caption{(color online) Limits on $\Lambda$ versus $1/\beta = M/M_{Pl}$ for the chameleon model discussed in the text.  The 95\% CL exclusion limits from this search are denoted by the dark (gray) region.  Recent constraints from atom interferometry are shown by the light (blue) region~\cite{muller15,Elder16}.  The horizontal line indicates $\Lambda = 2.4$~meV.  Limits from neutron interferometry~\cite{Li16,Lemmel15,Jenke14} and from the E\"{o}t-Wash torsion balance experiment~\cite{adelberger07,Adelberger06,Upadhye12} are denoted by the hatched regions. These limits are shown only in the restricted regions of parameter space considered in Refs~\cite{Li16} and \cite{Upadhye12}.}
\label{fig:limits}
\end{figure}

The resulting limits on $1/\beta = M/M_{Pl}$ are shown in Fig.~\ref{fig:limits} and compared to existing limits on chameleon interactions.
Due to the self-screening of the microsphere at large values of $\beta$, these results are not able to constrain forces arising from chameleons for $\Lambda = 2.4$~meV given current backgrounds.  However, at values of $\Lambda > 4.9$~meV, the self-screening is reduced, and these data are able to constrain chameleon interactions. These bounds are within a factor of 3 of the best existing constraints from atom interferometry using an entirely independent technique.

The analysis presented here constrains screened interactions that would produce a force between the cantilever and the microsphere greater than 0.1~fN at separations greater than 20~$\mu$m. This search is limited by backgrounds from fixed dipole moments in the microspheres coupling to electric fields caused by contact potentials. One method for reducing such backgrounds is to spin the microspheres by applying an optical~\cite{bishop2004} or electrostatic torque~\cite{simpson2002}. It might be possible to anneal the microspheres in situ~\cite{Millen14} to increase the rate at which separated charges within the microspheres recombine. Finally, commercially available microspheres made from different materials might have smaller permanent dipole moments. Some combination of these techniques may be used in the future to enhance the sensitivity reached here. 

These results provide the first search for interactions below the dark energy length scale using isolated mesoscopic objects separated by mesoscopic distances without an intervening electrostatic shield. This experimental technique is complementary to previous searches and could be sensitive to interactions that have evaded detection to date. The determination of the electric field near the cantilever and measurement of the interaction of electrically neutral silica microspheres with these fields provides important constraints on the expected backgrounds for future searches using similar methods. Future work will feature a search optimized for unscreened Yukawa interactions.  

We would like to thank P.~Brax for useful discussions during the early stages of this work.  This work is supported in part by NSF grant PHY1502156.  ADR is supported by an ARCS Foundation Stanford Graduate Fellowship. Nanofabrication was performed at the Stanford Nanofabrication Facility.    

\bibliography{chameleons}

\begin{thebibliography}{47}%
\makeatletter
\providecommand \@ifxundefined [1]{%
 \@ifx{#1\undefined}
}%
\providecommand \@ifnum [1]{%
 \ifnum #1\expandafter \@firstoftwo
 \else \expandafter \@secondoftwo
 \fi
}%
\providecommand \@ifx [1]{%
 \ifx #1\expandafter \@firstoftwo
 \else \expandafter \@secondoftwo
 \fi
}%
\providecommand \natexlab [1]{#1}%
\providecommand \enquote  [1]{``#1''}%
\providecommand \bibnamefont  [1]{#1}%
\providecommand \bibfnamefont [1]{#1}%
\providecommand \citenamefont [1]{#1}%
\providecommand \href@noop [0]{\@secondoftwo}%
\providecommand \href [0]{\begingroup \@sanitize@url \@href}%
\providecommand \@href[1]{\@@startlink{#1}\@@href}%
\providecommand \@@href[1]{\endgroup#1\@@endlink}%
\providecommand \@sanitize@url [0]{\catcode `\\12\catcode `\$12\catcode
  `\&12\catcode `\#12\catcode `\^12\catcode `\_12\catcode `\%12\relax}%
\providecommand \@@startlink[1]{}%
\providecommand \@@endlink[0]{}%
\providecommand \url  [0]{\begingroup\@sanitize@url \@url }%
\providecommand \@url [1]{\endgroup\@href {#1}{\urlprefix }}%
\providecommand \urlprefix  [0]{URL }%
\providecommand \Eprint [0]{\href }%
\providecommand \doibase [0]{http://dx.doi.org/}%
\providecommand \selectlanguage [0]{\@gobble}%
\providecommand \bibinfo  [0]{\@secondoftwo}%
\providecommand \bibfield  [0]{\@secondoftwo}%
\providecommand \translation [1]{[#1]}%
\providecommand \BibitemOpen [0]{}%
\providecommand \bibitemStop [0]{}%
\providecommand \bibitemNoStop [0]{.\EOS\space}%
\providecommand \EOS [0]{\spacefactor3000\relax}%
\providecommand \BibitemShut  [1]{\csname bibitem#1\endcsname}%
\let\auto@bib@innerbib\@empty
%</preamble>
\bibitem [{\citenamefont {Riess}\ \emph {et~al.}(1998)\citenamefont {Riess}
  \emph {et~al.}}]{Riess98}%
  \BibitemOpen
  \bibfield  {author} {\bibinfo {author} {\bibfnamefont {A.~G.}\ \bibnamefont
  {Riess}} \emph {et~al.} (\bibinfo {collaboration} {Supernova Search Team}),\
  }\href {\doibase 10.1086/300499} {\bibfield  {journal} {\bibinfo  {journal}
  {Astron. J.}\ }\textbf {\bibinfo {volume} {116}},\ \bibinfo {pages} {1009}
  (\bibinfo {year} {1998})},\ \Eprint {http://arxiv.org/abs/astro-ph/9805201}
  {arXiv:astro-ph/9805201} \BibitemShut {NoStop}%
%%CITATION = ASTRO-PH/9805201;%%
\bibitem [{\citenamefont {Perlmutter}\ \emph {et~al.}(1999)\citenamefont
  {Perlmutter} \emph {et~al.}}]{Perlmutter98}%
  \BibitemOpen
  \bibfield  {author} {\bibinfo {author} {\bibfnamefont {S.}~\bibnamefont
  {Perlmutter}} \emph {et~al.} (\bibinfo {collaboration} {Supernova Cosmology
  Project}),\ }\href {\doibase 10.1086/307221} {\bibfield  {journal} {\bibinfo
  {journal} {Astrophys. J.}\ }\textbf {\bibinfo {volume} {517}},\ \bibinfo
  {pages} {565} (\bibinfo {year} {1999})},\ \Eprint
  {http://arxiv.org/abs/astro-ph/9812133} {arXiv:astro-ph/9812133} \BibitemShut
  {NoStop}%
%%CITATION = ASTRO-PH/9812133;%%
\bibitem [{\citenamefont {Ade}\ \emph {et~al.}(2015)\citenamefont {Ade} \emph
  {et~al.}}]{Ade15}%
  \BibitemOpen
  \bibfield  {author} {\bibinfo {author} {\bibfnamefont {P.~A.~R.}\
  \bibnamefont {Ade}} \emph {et~al.} (\bibinfo {collaboration} {Planck}),\
  }\href@noop {} {\  (\bibinfo {year} {2015})},\ \Eprint
  {http://arxiv.org/abs/1502.01589} {arXiv:1502.01589 [astro-ph.CO]}
  \BibitemShut {NoStop}%
%%CITATION = ARXIV:1502.01589;%%
\bibitem [{\citenamefont {Olive}\ \emph {et~al.}(2014)\citenamefont {Olive}
  \emph {et~al.}}]{Agashe2014}%
  \BibitemOpen
  \bibfield  {author} {\bibinfo {author} {\bibfnamefont {K.~A.}\ \bibnamefont
  {Olive}} \emph {et~al.} (\bibinfo {collaboration} {Particle Data Group}),\
  }\href {\doibase 10.1088/1674-1137/38/9/090001} {\bibfield  {journal}
  {\bibinfo  {journal} {Chin. Phys. C}\ }\textbf {\bibinfo {volume} {38}},\
  \bibinfo {pages} {090001} (\bibinfo {year} {2014})}\BibitemShut {NoStop}%
%%CITATION = CHPHD,C38,090001;%%
\bibitem [{\citenamefont {Jain}\ and\ \citenamefont {Khoury}(2010)}]{Jain10}%
  \BibitemOpen
  \bibfield  {author} {\bibinfo {author} {\bibfnamefont {B.}~\bibnamefont
  {Jain}}\ and\ \bibinfo {author} {\bibfnamefont {J.}~\bibnamefont {Khoury}},\
  }\href {\doibase 10.1016/j.aop.2010.04.002} {\bibfield  {journal} {\bibinfo
  {journal} {Annals Phys.}\ }\textbf {\bibinfo {volume} {325}},\ \bibinfo
  {pages} {1479} (\bibinfo {year} {2010})},\ \Eprint
  {http://arxiv.org/abs/1004.3294} {arXiv:1004.3294 [astro-ph.CO]} \BibitemShut
  {NoStop}%
%%CITATION = ARXIV:1004.3294;%%
\bibitem [{\citenamefont {Joyce}\ \emph {et~al.}(2015)\citenamefont {Joyce},
  \citenamefont {Jain}, \citenamefont {Khoury},\ and\ \citenamefont
  {Trodden}}]{Joyce14}%
  \BibitemOpen
  \bibfield  {author} {\bibinfo {author} {\bibfnamefont {A.}~\bibnamefont
  {Joyce}}, \bibinfo {author} {\bibfnamefont {B.}~\bibnamefont {Jain}},
  \bibinfo {author} {\bibfnamefont {J.}~\bibnamefont {Khoury}}, \ and\ \bibinfo
  {author} {\bibfnamefont {M.}~\bibnamefont {Trodden}},\ }\href {\doibase
  10.1016/j.physrep.2014.12.002} {\bibfield  {journal} {\bibinfo  {journal}
  {Phys. Rept.}\ }\textbf {\bibinfo {volume} {568}},\ \bibinfo {pages} {1}
  (\bibinfo {year} {2015})},\ \Eprint {http://arxiv.org/abs/1407.0059}
  {arXiv:1407.0059 [astro-ph.CO]} \BibitemShut {NoStop}%
%%CITATION = ARXIV:1407.0059;%%
\bibitem [{\citenamefont {{Beane}}(1997)}]{Beane97}%
  \BibitemOpen
  \bibfield  {author} {\bibinfo {author} {\bibfnamefont {S.~R.}\ \bibnamefont
  {{Beane}}},\ }\href {\doibase 10.1023/A:1018843607120} {\bibfield  {journal}
  {\bibinfo  {journal} {Gen. Rel. Grav.}\ }\textbf {\bibinfo {volume} {29}},\
  \bibinfo {pages} {945} (\bibinfo {year} {1997})},\ \Eprint
  {http://arxiv.org/abs/hep-ph/9702419} {arXiv:hep-ph/9702419} \BibitemShut
  {NoStop}%
\bibitem [{\citenamefont {Kapner}\ \emph {et~al.}(2007)\citenamefont {Kapner},
  \citenamefont {Cook}, \citenamefont {Adelberger}, \citenamefont {Gundlach},
  \citenamefont {Heckel}, \citenamefont {Hoyle},\ and\ \citenamefont
  {Swanson}}]{adelberger07}%
  \BibitemOpen
  \bibfield  {author} {\bibinfo {author} {\bibfnamefont {D.~J.}\ \bibnamefont
  {Kapner}}, \bibinfo {author} {\bibfnamefont {T.~S.}\ \bibnamefont {Cook}},
  \bibinfo {author} {\bibfnamefont {E.~G.}\ \bibnamefont {Adelberger}},
  \bibinfo {author} {\bibfnamefont {J.~H.}\ \bibnamefont {Gundlach}}, \bibinfo
  {author} {\bibfnamefont {B.~R.}\ \bibnamefont {Heckel}}, \bibinfo {author}
  {\bibfnamefont {C.~D.}\ \bibnamefont {Hoyle}}, \ and\ \bibinfo {author}
  {\bibfnamefont {H.~E.}\ \bibnamefont {Swanson}},\ }\href {\doibase
  10.1103/PhysRevLett.98.021101} {\bibfield  {journal} {\bibinfo  {journal}
  {Phys. Rev. Lett.}\ }\textbf {\bibinfo {volume} {98}},\ \bibinfo {pages}
  {021101} (\bibinfo {year} {2007})},\ \Eprint {http://arxiv.org/abs/0611184}
  {arXiv:0611184 [hep-ph]} \BibitemShut {NoStop}%
\bibitem [{\citenamefont {Mota}\ and\ \citenamefont {Shaw}(2006)}]{Mota06}%
  \BibitemOpen
  \bibfield  {author} {\bibinfo {author} {\bibfnamefont {D.~F.}\ \bibnamefont
  {Mota}}\ and\ \bibinfo {author} {\bibfnamefont {D.~J.}\ \bibnamefont
  {Shaw}},\ }\href {\doibase 10.1103/PhysRevLett.97.151102} {\bibfield
  {journal} {\bibinfo  {journal} {Phys. Rev. Lett.}\ }\textbf {\bibinfo
  {volume} {97}},\ \bibinfo {pages} {151102} (\bibinfo {year} {2006})},\
  \Eprint {http://arxiv.org/abs/hep-ph/0606204} {arXiv:hep-ph/0606204}
  \BibitemShut {NoStop}%
%%CITATION = HEP-PH/0606204;%%
\bibitem [{\citenamefont {Adelberger}\ \emph {et~al.}(2009)\citenamefont
  {Adelberger}, \citenamefont {Gundlach}, \citenamefont {Heckel}, \citenamefont
  {Hoedl},\ and\ \citenamefont {Schlamminger}}]{Adelberger09}%
  \BibitemOpen
  \bibfield  {author} {\bibinfo {author} {\bibfnamefont {E.~G.}\ \bibnamefont
  {Adelberger}}, \bibinfo {author} {\bibfnamefont {J.~H.}\ \bibnamefont
  {Gundlach}}, \bibinfo {author} {\bibfnamefont {B.~R.}\ \bibnamefont
  {Heckel}}, \bibinfo {author} {\bibfnamefont {S.}~\bibnamefont {Hoedl}}, \
  and\ \bibinfo {author} {\bibfnamefont {S.}~\bibnamefont {Schlamminger}},\
  }\href {\doibase 10.1016/j.ppnp.2008.08.002} {\bibfield  {journal} {\bibinfo
  {journal} {Prog. Part. Nucl. Phys.}\ }\textbf {\bibinfo {volume} {62}},\
  \bibinfo {pages} {102} (\bibinfo {year} {2009})}\BibitemShut {NoStop}%
%%CITATION = PPNPD,62,102;%%
\bibitem [{\citenamefont {Geraci}\ \emph {et~al.}(2008)\citenamefont {Geraci},
  \citenamefont {Smullin}, \citenamefont {Weld}, \citenamefont {Chiaverini},\
  and\ \citenamefont {Kapitulnik}}]{Geraci08}%
  \BibitemOpen
  \bibfield  {author} {\bibinfo {author} {\bibfnamefont {A.~A.}\ \bibnamefont
  {Geraci}}, \bibinfo {author} {\bibfnamefont {S.~J.}\ \bibnamefont {Smullin}},
  \bibinfo {author} {\bibfnamefont {D.~M.}\ \bibnamefont {Weld}}, \bibinfo
  {author} {\bibfnamefont {J.}~\bibnamefont {Chiaverini}}, \ and\ \bibinfo
  {author} {\bibfnamefont {A.}~\bibnamefont {Kapitulnik}},\ }\href {\doibase
  10.1103/PhysRevD.78.022002} {\bibfield  {journal} {\bibinfo  {journal} {Phys.
  Rev. D}\ }\textbf {\bibinfo {volume} {78}},\ \bibinfo {pages} {022002}
  (\bibinfo {year} {2008})},\ \Eprint {http://arxiv.org/abs/0802.2350}
  {arXiv:0802.2350 [hep-ex]} \BibitemShut {NoStop}%
\bibitem [{\citenamefont {Sushkov}\ \emph {et~al.}(2011)\citenamefont
  {Sushkov}, \citenamefont {Kim}, \citenamefont {Dalvit},\ and\ \citenamefont
  {Lamoreaux}}]{Sushkov11}%
  \BibitemOpen
  \bibfield  {author} {\bibinfo {author} {\bibfnamefont {A.~O.}\ \bibnamefont
  {Sushkov}}, \bibinfo {author} {\bibfnamefont {W.~J.}\ \bibnamefont {Kim}},
  \bibinfo {author} {\bibfnamefont {D.~A.~R.}\ \bibnamefont {Dalvit}}, \ and\
  \bibinfo {author} {\bibfnamefont {S.~K.}\ \bibnamefont {Lamoreaux}},\ }\href
  {\doibase 10.1103/PhysRevLett.107.171101} {\bibfield  {journal} {\bibinfo
  {journal} {Phys. Rev. Lett.}\ }\textbf {\bibinfo {volume} {107}},\ \bibinfo
  {pages} {171101} (\bibinfo {year} {2011})},\ \Eprint
  {http://arxiv.org/abs/1108.2547} {arXiv:1108.2547 [quant-ph]} \BibitemShut
  {NoStop}%
\bibitem [{\citenamefont {Tan}\ \emph {et~al.}(2016)\citenamefont {Tan},
  \citenamefont {Yang}, \citenamefont {Shao}, \citenamefont {Li}, \citenamefont
  {Du}, \citenamefont {Zhan}, \citenamefont {Wang}, \citenamefont {Luo},
  \citenamefont {Tu},\ and\ \citenamefont {Luo}}]{Wen16}%
  \BibitemOpen
  \bibfield  {author} {\bibinfo {author} {\bibfnamefont {W.-H.}\ \bibnamefont
  {Tan}}, \bibinfo {author} {\bibfnamefont {S.-Q.}\ \bibnamefont {Yang}},
  \bibinfo {author} {\bibfnamefont {C.-G.}\ \bibnamefont {Shao}}, \bibinfo
  {author} {\bibfnamefont {J.}~\bibnamefont {Li}}, \bibinfo {author}
  {\bibfnamefont {A.-B.}\ \bibnamefont {Du}}, \bibinfo {author} {\bibfnamefont
  {B.-F.}\ \bibnamefont {Zhan}}, \bibinfo {author} {\bibfnamefont {Q.-L.}\
  \bibnamefont {Wang}}, \bibinfo {author} {\bibfnamefont {P.-S.}\ \bibnamefont
  {Luo}}, \bibinfo {author} {\bibfnamefont {L.-C.}\ \bibnamefont {Tu}}, \ and\
  \bibinfo {author} {\bibfnamefont {J.}~\bibnamefont {Luo}},\ }\href {\doibase
  10.1103/PhysRevLett.116.131101} {\bibfield  {journal} {\bibinfo  {journal}
  {Phys. Rev. Lett.}\ }\textbf {\bibinfo {volume} {116}},\ \bibinfo {pages}
  {131101} (\bibinfo {year} {2016})}\BibitemShut {NoStop}%
\bibitem [{\citenamefont {Upadhye}(2012)}]{Upadhye12}%
  \BibitemOpen
  \bibfield  {author} {\bibinfo {author} {\bibfnamefont {A.}~\bibnamefont
  {Upadhye}},\ }\href {\doibase 10.1103/PhysRevD.86.102003} {\bibfield
  {journal} {\bibinfo  {journal} {Phys. Rev. D}\ }\textbf {\bibinfo {volume}
  {86}},\ \bibinfo {pages} {102003} (\bibinfo {year} {2012})},\ \Eprint
  {http://arxiv.org/abs/1209.0211} {arXiv:1209.0211 [hep-ph]} \BibitemShut
  {NoStop}%
%%CITATION = ARXIV:1209.0211;%%
\bibitem [{\citenamefont {Burrage}\ \emph {et~al.}(2015)\citenamefont
  {Burrage}, \citenamefont {Copeland},\ and\ \citenamefont
  {Hinds}}]{Burrage15}%
  \BibitemOpen
  \bibfield  {author} {\bibinfo {author} {\bibfnamefont {C.}~\bibnamefont
  {Burrage}}, \bibinfo {author} {\bibfnamefont {E.~J.}\ \bibnamefont
  {Copeland}}, \ and\ \bibinfo {author} {\bibfnamefont {E.~A.}\ \bibnamefont
  {Hinds}},\ }\href {\doibase 10.1088/1475-7516/2015/03/042} {\bibfield
  {journal} {\bibinfo  {journal} {JCAP}\ }\textbf {\bibinfo {volume} {1503}},\
  \bibinfo {pages} {042} (\bibinfo {year} {2015})},\ \Eprint
  {http://arxiv.org/abs/1408.1409} {arXiv:1408.1409 [astro-ph.CO]} \BibitemShut
  {NoStop}%
%%CITATION = ARXIV:1408.1409;%%
\bibitem [{\citenamefont {Hamilton}\ \emph {et~al.}(2015)\citenamefont
  {Hamilton}, \citenamefont {Jaffe}, \citenamefont {Haslinger}, \citenamefont
  {Simmons}, \citenamefont {M{\"u}ller},\ and\ \citenamefont
  {Khoury}}]{muller15}%
  \BibitemOpen
  \bibfield  {author} {\bibinfo {author} {\bibfnamefont {P.}~\bibnamefont
  {Hamilton}}, \bibinfo {author} {\bibfnamefont {M.}~\bibnamefont {Jaffe}},
  \bibinfo {author} {\bibfnamefont {P.}~\bibnamefont {Haslinger}}, \bibinfo
  {author} {\bibfnamefont {Q.}~\bibnamefont {Simmons}}, \bibinfo {author}
  {\bibfnamefont {H.}~\bibnamefont {M{\"u}ller}}, \ and\ \bibinfo {author}
  {\bibfnamefont {J.}~\bibnamefont {Khoury}},\ }\href@noop {} {\bibfield
  {journal} {\bibinfo  {journal} {Science}\ }\textbf {\bibinfo {volume}
  {349}},\ \bibinfo {pages} {849–851} (\bibinfo {year} {2015})},\ \Eprint
  {http://arxiv.org/abs/1502.03888} {arXiv:1502.03888 [physics.atom-ph]}
  \BibitemShut {NoStop}%
\bibitem [{\citenamefont {Elder}\ \emph {et~al.}(2016)\citenamefont {Elder},
  \citenamefont {Khoury}, \citenamefont {Haslinger}, \citenamefont {Jaffe},
  \citenamefont {M{\"u}ller},\ and\ \citenamefont {Hamilton}}]{Elder16}%
  \BibitemOpen
  \bibfield  {author} {\bibinfo {author} {\bibfnamefont {B.}~\bibnamefont
  {Elder}}, \bibinfo {author} {\bibfnamefont {J.}~\bibnamefont {Khoury}},
  \bibinfo {author} {\bibfnamefont {P.}~\bibnamefont {Haslinger}}, \bibinfo
  {author} {\bibfnamefont {M.}~\bibnamefont {Jaffe}}, \bibinfo {author}
  {\bibfnamefont {H.}~\bibnamefont {M{\"u}ller}}, \ and\ \bibinfo {author}
  {\bibfnamefont {P.}~\bibnamefont {Hamilton}},\ }\href@noop {} {\  (\bibinfo
  {year} {2016})},\ \Eprint {http://arxiv.org/abs/1603.06587} {arXiv:1603.06587
  [astro-ph.CO]} \BibitemShut {NoStop}%
%%CITATION = ARXIV:1603.06587;%%
\bibitem [{\citenamefont {Li}\ \emph {et~al.}(2016)\citenamefont {Li},
  \citenamefont {Arif}, \citenamefont {Cory}, \citenamefont {Haun},
  \citenamefont {Heacock}, \citenamefont {Huber}, \citenamefont {Nsofini},
  \citenamefont {Pushin}, \citenamefont {Saggu}, \citenamefont {Sarenac},
  \citenamefont {Shahi}, \citenamefont {Skavysh}, \citenamefont {Snow},\ and\
  \citenamefont {Young}}]{Li16}%
  \BibitemOpen
  \bibfield  {author} {\bibinfo {author} {\bibfnamefont {K.}~\bibnamefont
  {Li}}, \bibinfo {author} {\bibfnamefont {M.}~\bibnamefont {Arif}}, \bibinfo
  {author} {\bibfnamefont {D.~G.}\ \bibnamefont {Cory}}, \bibinfo {author}
  {\bibfnamefont {R.}~\bibnamefont {Haun}}, \bibinfo {author} {\bibfnamefont
  {B.}~\bibnamefont {Heacock}}, \bibinfo {author} {\bibfnamefont {M.~G.}\
  \bibnamefont {Huber}}, \bibinfo {author} {\bibfnamefont {J.}~\bibnamefont
  {Nsofini}}, \bibinfo {author} {\bibfnamefont {D.~A.}\ \bibnamefont {Pushin}},
  \bibinfo {author} {\bibfnamefont {P.}~\bibnamefont {Saggu}}, \bibinfo
  {author} {\bibfnamefont {D.}~\bibnamefont {Sarenac}}, \bibinfo {author}
  {\bibfnamefont {C.~B.}\ \bibnamefont {Shahi}}, \bibinfo {author}
  {\bibfnamefont {V.}~\bibnamefont {Skavysh}}, \bibinfo {author} {\bibfnamefont
  {W.~M.}\ \bibnamefont {Snow}}, \ and\ \bibinfo {author} {\bibfnamefont
  {A.~R.}\ \bibnamefont {Young}} (\bibinfo {collaboration} {The INDEX
  Collaboration}),\ }\href {\doibase 10.1103/PhysRevD.93.062001} {\bibfield
  {journal} {\bibinfo  {journal} {Phys. Rev. D}\ }\textbf {\bibinfo {volume}
  {93}},\ \bibinfo {pages} {062001} (\bibinfo {year} {2016})},\ \Eprint
  {http://arxiv.org/abs/1601.06897} {arXiv:1601.06897 [astro-ph.CO]}
  \BibitemShut {NoStop}%
%%CITATION = ARXIV:1601.06897;%%
\bibitem [{\citenamefont {Lemmel}\ \emph {et~al.}(2015)\citenamefont {Lemmel},
  \citenamefont {Brax}, \citenamefont {Ivanov}, \citenamefont {Jenke},
  \citenamefont {Pignol}, \citenamefont {Pitschmann}, \citenamefont {Potocar},
  \citenamefont {Wellenzohn}, \citenamefont {Zawisky},\ and\ \citenamefont
  {Abele}}]{Lemmel15}%
  \BibitemOpen
  \bibfield  {author} {\bibinfo {author} {\bibfnamefont {H.}~\bibnamefont
  {Lemmel}}, \bibinfo {author} {\bibfnamefont {P.}~\bibnamefont {Brax}},
  \bibinfo {author} {\bibfnamefont {A.~N.}\ \bibnamefont {Ivanov}}, \bibinfo
  {author} {\bibfnamefont {T.}~\bibnamefont {Jenke}}, \bibinfo {author}
  {\bibfnamefont {G.}~\bibnamefont {Pignol}}, \bibinfo {author} {\bibfnamefont
  {M.}~\bibnamefont {Pitschmann}}, \bibinfo {author} {\bibfnamefont
  {T.}~\bibnamefont {Potocar}}, \bibinfo {author} {\bibfnamefont
  {M.}~\bibnamefont {Wellenzohn}}, \bibinfo {author} {\bibfnamefont
  {M.}~\bibnamefont {Zawisky}}, \ and\ \bibinfo {author} {\bibfnamefont
  {H.}~\bibnamefont {Abele}},\ }\href {\doibase 10.1016/j.physletb.2015.02.063}
  {\bibfield  {journal} {\bibinfo  {journal} {Phys. Lett. B}\ }\textbf
  {\bibinfo {volume} {743}},\ \bibinfo {pages} {310} (\bibinfo {year}
  {2015})},\ \Eprint {http://arxiv.org/abs/1502.06023} {arXiv:1502.06023
  [hep-ph]} \BibitemShut {NoStop}%
%%CITATION = ARXIV:1502.06023;%%
\bibitem [{\citenamefont {Jenke}\ \emph {et~al.}(2014)\citenamefont {Jenke},
  \citenamefont {Cronenberg}, \citenamefont {Burgd\"orfer}, \citenamefont
  {Chizhova}, \citenamefont {Geltenbort}, \citenamefont {Ivanov}, \citenamefont
  {Lauer}, \citenamefont {Lins}, \citenamefont {Rotter}, \citenamefont {Saul},
  \citenamefont {Schmidt},\ and\ \citenamefont {Abele}}]{Jenke14}%
  \BibitemOpen
  \bibfield  {author} {\bibinfo {author} {\bibfnamefont {T.}~\bibnamefont
  {Jenke}}, \bibinfo {author} {\bibfnamefont {G.}~\bibnamefont {Cronenberg}},
  \bibinfo {author} {\bibfnamefont {J.}~\bibnamefont {Burgd\"orfer}}, \bibinfo
  {author} {\bibfnamefont {L.~A.}\ \bibnamefont {Chizhova}}, \bibinfo {author}
  {\bibfnamefont {P.}~\bibnamefont {Geltenbort}}, \bibinfo {author}
  {\bibfnamefont {A.~N.}\ \bibnamefont {Ivanov}}, \bibinfo {author}
  {\bibfnamefont {T.}~\bibnamefont {Lauer}}, \bibinfo {author} {\bibfnamefont
  {T.}~\bibnamefont {Lins}}, \bibinfo {author} {\bibfnamefont {S.}~\bibnamefont
  {Rotter}}, \bibinfo {author} {\bibfnamefont {H.}~\bibnamefont {Saul}},
  \bibinfo {author} {\bibfnamefont {U.}~\bibnamefont {Schmidt}}, \ and\
  \bibinfo {author} {\bibfnamefont {H.}~\bibnamefont {Abele}},\ }\href
  {\doibase 10.1103/PhysRevLett.112.151105} {\bibfield  {journal} {\bibinfo
  {journal} {Phys. Rev. Lett.}\ }\textbf {\bibinfo {volume} {112}},\ \bibinfo
  {pages} {151105} (\bibinfo {year} {2014})},\ \Eprint
  {http://arxiv.org/abs/1404.4099} {arXiv:1404.4099 [gr-qc]} \BibitemShut
  {NoStop}%
%%CITATION = ARXIV:1404.4099;%%
\bibitem [{\citenamefont {Khoury}\ and\ \citenamefont
  {Weltman}(2004{\natexlab{a}})}]{Khoury03a}%
  \BibitemOpen
  \bibfield  {author} {\bibinfo {author} {\bibfnamefont {J.}~\bibnamefont
  {Khoury}}\ and\ \bibinfo {author} {\bibfnamefont {A.}~\bibnamefont
  {Weltman}},\ }\href {\doibase 10.1103/PhysRevLett.93.171104} {\bibfield
  {journal} {\bibinfo  {journal} {Phys. Rev. Lett.}\ }\textbf {\bibinfo
  {volume} {93}},\ \bibinfo {pages} {171104} (\bibinfo {year}
  {2004}{\natexlab{a}})},\ \Eprint {http://arxiv.org/abs/0309300}
  {arXiv:0309300 [astro-ph]} \BibitemShut {NoStop}%
%%CITATION = ASTRO-PH/0309300;%%
\bibitem [{\citenamefont {Khoury}\ and\ \citenamefont
  {Weltman}(2004{\natexlab{b}})}]{Khoury03b}%
  \BibitemOpen
  \bibfield  {author} {\bibinfo {author} {\bibfnamefont {J.}~\bibnamefont
  {Khoury}}\ and\ \bibinfo {author} {\bibfnamefont {A.}~\bibnamefont
  {Weltman}},\ }\href {\doibase 10.1103/PhysRevD.69.044026} {\bibfield
  {journal} {\bibinfo  {journal} {Phys. Rev. D}\ }\textbf {\bibinfo {volume}
  {69}},\ \bibinfo {pages} {044026} (\bibinfo {year} {2004}{\natexlab{b}})},\
  \Eprint {http://arxiv.org/abs/0309411} {arXiv:0309411 [astro-ph]}
  \BibitemShut {NoStop}%
%%CITATION = ASTRO-PH/0309411;%%
\bibitem [{\citenamefont {{Brax}}\ \emph {et~al.}(2004)\citenamefont {{Brax}},
  \citenamefont {{van de Bruck}}, \citenamefont {{Davis}}, \citenamefont
  {{Khoury}},\ and\ \citenamefont {{Weltman}}}]{brax04}%
  \BibitemOpen
  \bibfield  {author} {\bibinfo {author} {\bibfnamefont {P.}~\bibnamefont
  {{Brax}}}, \bibinfo {author} {\bibfnamefont {C.}~\bibnamefont {{van de
  Bruck}}}, \bibinfo {author} {\bibfnamefont {A.~C.}\ \bibnamefont {{Davis}}},
  \bibinfo {author} {\bibfnamefont {J.}~\bibnamefont {{Khoury}}}, \ and\
  \bibinfo {author} {\bibfnamefont {A.}~\bibnamefont {{Weltman}}},\ }\href
  {\doibase 10.1063/1.1835177} {\bibfield  {journal} {\bibinfo  {journal} {AIP
  Conf. Ser.}\ }\textbf {\bibinfo {volume} {736}},\ \bibinfo {pages} {105}
  (\bibinfo {year} {2004})},\ \Eprint {http://arxiv.org/abs/astro-ph/0410103}
  {arXiv:astro-ph/0410103} \BibitemShut {NoStop}%
\bibitem [{\citenamefont {{Chang}}\ \emph {et~al.}(2010)\citenamefont
  {{Chang}}, \citenamefont {{Regal}}, \citenamefont {{Papp}}, \citenamefont
  {{Wilson}}, \citenamefont {{Ye}}, \citenamefont {{Painter}}, \citenamefont
  {{Kimble}},\ and\ \citenamefont {{Zoller}}}]{Chang10}%
  \BibitemOpen
  \bibfield  {author} {\bibinfo {author} {\bibfnamefont {D.~E.}\ \bibnamefont
  {{Chang}}}, \bibinfo {author} {\bibfnamefont {C.~A.}\ \bibnamefont
  {{Regal}}}, \bibinfo {author} {\bibfnamefont {S.~B.}\ \bibnamefont {{Papp}}},
  \bibinfo {author} {\bibfnamefont {D.~J.}\ \bibnamefont {{Wilson}}}, \bibinfo
  {author} {\bibfnamefont {J.}~\bibnamefont {{Ye}}}, \bibinfo {author}
  {\bibfnamefont {O.}~\bibnamefont {{Painter}}}, \bibinfo {author}
  {\bibfnamefont {H.~J.}\ \bibnamefont {{Kimble}}}, \ and\ \bibinfo {author}
  {\bibfnamefont {P.}~\bibnamefont {{Zoller}}},\ }\href {\doibase
  10.1073/pnas.0912969107} {\bibfield  {journal} {\bibinfo  {journal} {Proc.
  Natl. Acad. Sci.}\ }\textbf {\bibinfo {volume} {107}},\ \bibinfo {pages}
  {1005} (\bibinfo {year} {2010})},\ \Eprint {http://arxiv.org/abs/0909.1548}
  {arXiv:0909.1548 [quant-ph]} \BibitemShut {NoStop}%
\bibitem [{\citenamefont {{Romero-Isart}}\ \emph {et~al.}(2010)\citenamefont
  {{Romero-Isart}}, \citenamefont {{Juan}}, \citenamefont {{Quidant}},\ and\
  \citenamefont {{Cirac}}}]{Romero10}%
  \BibitemOpen
  \bibfield  {author} {\bibinfo {author} {\bibfnamefont {O.}~\bibnamefont
  {{Romero-Isart}}}, \bibinfo {author} {\bibfnamefont {M.~L.}\ \bibnamefont
  {{Juan}}}, \bibinfo {author} {\bibfnamefont {R.}~\bibnamefont {{Quidant}}}, \
  and\ \bibinfo {author} {\bibfnamefont {J.~I.}\ \bibnamefont {{Cirac}}},\
  }\href {\doibase 10.1088/1367-2630/12/3/033015} {\bibfield  {journal}
  {\bibinfo  {journal} {New J. of Phys.}\ }\textbf {\bibinfo {volume} {12}},\
  \bibinfo {eid} {033015} (\bibinfo {year} {2010})},\ \Eprint
  {http://arxiv.org/abs/0909.1469} {arXiv:0909.1469 [quant-ph]} \BibitemShut
  {NoStop}%
\bibitem [{\citenamefont {Yin}\ \emph {et~al.}(2013)\citenamefont {Yin},
  \citenamefont {Geraci},\ and\ \citenamefont {Li}}]{Yin13}%
  \BibitemOpen
  \bibfield  {author} {\bibinfo {author} {\bibfnamefont {Z.-Q.}\ \bibnamefont
  {Yin}}, \bibinfo {author} {\bibfnamefont {A.~A.}\ \bibnamefont {Geraci}}, \
  and\ \bibinfo {author} {\bibfnamefont {T.}~\bibnamefont {Li}},\ }\href
  {\doibase 10.1142/S0217979213300181} {\bibfield  {journal} {\bibinfo
  {journal} {Int. J. Mod. Phys.}\ }\textbf {\bibinfo {volume} {B27}},\ \bibinfo
  {pages} {1330018} (\bibinfo {year} {2013})},\ \Eprint
  {http://arxiv.org/abs/1308.4503} {arXiv:1308.4503 [quant-ph]} \BibitemShut
  {NoStop}%
%%CITATION = ARXIV:1308.4503;%%
\bibitem [{\citenamefont {Li}\ \emph {et~al.}(2011)\citenamefont {Li},
  \citenamefont {Kheifets},\ and\ \citenamefont {Raizen}}]{Li11}%
  \BibitemOpen
  \bibfield  {author} {\bibinfo {author} {\bibfnamefont {T.}~\bibnamefont
  {Li}}, \bibinfo {author} {\bibfnamefont {S.}~\bibnamefont {Kheifets}}, \ and\
  \bibinfo {author} {\bibfnamefont {M.~G.}\ \bibnamefont {Raizen}},\ }\href
  {\doibase 10.1038/nphys1952} {\bibfield  {journal} {\bibinfo  {journal}
  {Nature Phys.}\ }\textbf {\bibinfo {volume} {7}},\ \bibinfo {pages} {527}
  (\bibinfo {year} {2011})},\ \Eprint {http://arxiv.org/abs/1101.1283}
  {arXiv:1101.1283 [quant-ph]} \BibitemShut {NoStop}%
%%CITATION = ARXIV:1101.1283;%%
\bibitem [{\citenamefont {Gieseler}\ \emph {et~al.}(2012)\citenamefont
  {Gieseler}, \citenamefont {Deutsch}, \citenamefont {Quidant},\ and\
  \citenamefont {Novotny}}]{Gieseler12}%
  \BibitemOpen
  \bibfield  {author} {\bibinfo {author} {\bibfnamefont {J.}~\bibnamefont
  {Gieseler}}, \bibinfo {author} {\bibfnamefont {B.}~\bibnamefont {Deutsch}},
  \bibinfo {author} {\bibfnamefont {R.}~\bibnamefont {Quidant}}, \ and\
  \bibinfo {author} {\bibfnamefont {L.}~\bibnamefont {Novotny}},\ }\href
  {\doibase 10.1103/PhysRevLett.109.103603} {\bibfield  {journal} {\bibinfo
  {journal} {Phys. Rev. Lett.}\ }\textbf {\bibinfo {volume} {109}},\ \bibinfo
  {pages} {103603} (\bibinfo {year} {2012})},\ \Eprint
  {http://arxiv.org/abs/1202.6435} {arXiv:1202.6435 [cond-mat.mes-hall]}
  \BibitemShut {NoStop}%
\bibitem [{\citenamefont {{Kiesel}}\ \emph {et~al.}(2013)\citenamefont
  {{Kiesel}}, \citenamefont {{Blaser}}, \citenamefont {{Delic}}, \citenamefont
  {{Grass}}, \citenamefont {{Kaltenbaek}},\ and\ \citenamefont
  {{Aspelmeyer}}}]{Kiesel13}%
  \BibitemOpen
  \bibfield  {author} {\bibinfo {author} {\bibfnamefont {N.}~\bibnamefont
  {{Kiesel}}}, \bibinfo {author} {\bibfnamefont {F.}~\bibnamefont {{Blaser}}},
  \bibinfo {author} {\bibfnamefont {U.}~\bibnamefont {{Delic}}}, \bibinfo
  {author} {\bibfnamefont {D.}~\bibnamefont {{Grass}}}, \bibinfo {author}
  {\bibfnamefont {R.}~\bibnamefont {{Kaltenbaek}}}, \ and\ \bibinfo {author}
  {\bibfnamefont {M.}~\bibnamefont {{Aspelmeyer}}},\ }\href {\doibase
  10.1073/pnas.1309167110} {\bibfield  {journal} {\bibinfo  {journal} {Proc.
  Natl. Acad. Sci.}\ }\textbf {\bibinfo {volume} {110}},\ \bibinfo {pages}
  {14180} (\bibinfo {year} {2013})},\ \Eprint {http://arxiv.org/abs/1304.6679}
  {arXiv:1304.6679 [quant-ph]} \BibitemShut {NoStop}%
\bibitem [{\citenamefont {Ranjit}\ \emph {et~al.}(2015)\citenamefont {Ranjit},
  \citenamefont {Atherton}, \citenamefont {Stutz}, \citenamefont {Cunningham},\
  and\ \citenamefont {Geraci}}]{Ranjit15}%
  \BibitemOpen
  \bibfield  {author} {\bibinfo {author} {\bibfnamefont {G.}~\bibnamefont
  {Ranjit}}, \bibinfo {author} {\bibfnamefont {D.~P.}\ \bibnamefont
  {Atherton}}, \bibinfo {author} {\bibfnamefont {J.~H.}\ \bibnamefont {Stutz}},
  \bibinfo {author} {\bibfnamefont {M.}~\bibnamefont {Cunningham}}, \ and\
  \bibinfo {author} {\bibfnamefont {A.~A.}\ \bibnamefont {Geraci}},\ }\href
  {\doibase 10.1103/PhysRevA.91.051805} {\bibfield  {journal} {\bibinfo
  {journal} {Phys. Rev. A}\ }\textbf {\bibinfo {volume} {91}},\ \bibinfo
  {pages} {051805} (\bibinfo {year} {2015})},\ \Eprint
  {http://arxiv.org/abs/1503.08799} {arXiv:1503.08799 [physics.optics]}
  \BibitemShut {NoStop}%
\bibitem [{\citenamefont {Millen}\ \emph {et~al.}(2015)\citenamefont {Millen},
  \citenamefont {Fonseca}, \citenamefont {Mavrogordatos}, \citenamefont
  {Monteiro},\ and\ \citenamefont {Barker}}]{Millen15}%
  \BibitemOpen
  \bibfield  {author} {\bibinfo {author} {\bibfnamefont {J.}~\bibnamefont
  {Millen}}, \bibinfo {author} {\bibfnamefont {P.~Z.~G.}\ \bibnamefont
  {Fonseca}}, \bibinfo {author} {\bibfnamefont {T.}~\bibnamefont
  {Mavrogordatos}}, \bibinfo {author} {\bibfnamefont {T.~S.}\ \bibnamefont
  {Monteiro}}, \ and\ \bibinfo {author} {\bibfnamefont {P.~F.}\ \bibnamefont
  {Barker}},\ }\href {\doibase 10.1103/PhysRevLett.114.123602} {\bibfield
  {journal} {\bibinfo  {journal} {Phys. Rev. Lett.}\ }\textbf {\bibinfo
  {volume} {114}},\ \bibinfo {pages} {123602} (\bibinfo {year} {2015})},\
  \Eprint {http://arxiv.org/abs/1407.3595} {arXiv:1407.3595 [physics.optics]}
  \BibitemShut {NoStop}%
\bibitem [{\citenamefont {{Fonseca}}\ \emph {et~al.}(2015)\citenamefont
  {{Fonseca}}, \citenamefont {{Aranas}}, \citenamefont {{Millen}},
  \citenamefont {{Monteiro}},\ and\ \citenamefont {{Barker}}}]{Fonseca15}%
  \BibitemOpen
  \bibfield  {author} {\bibinfo {author} {\bibfnamefont {P.~Z.~G.}\
  \bibnamefont {{Fonseca}}}, \bibinfo {author} {\bibfnamefont {E.~B.}\
  \bibnamefont {{Aranas}}}, \bibinfo {author} {\bibfnamefont {J.}~\bibnamefont
  {{Millen}}}, \bibinfo {author} {\bibfnamefont {T.~S.}\ \bibnamefont
  {{Monteiro}}}, \ and\ \bibinfo {author} {\bibfnamefont {P.~F.}\ \bibnamefont
  {{Barker}}},\ }\href@noop {} {\  (\bibinfo {year} {2015})},\ \Eprint
  {http://arxiv.org/abs/1511.08482} {arXiv:1511.08482 [quant-ph]} \BibitemShut
  {NoStop}%
\bibitem [{\citenamefont {{Geraci}}\ \emph {et~al.}(2010)\citenamefont
  {{Geraci}}, \citenamefont {{Papp}},\ and\ \citenamefont
  {{Kitching}}}]{Geraci10}%
  \BibitemOpen
  \bibfield  {author} {\bibinfo {author} {\bibfnamefont {A.~A.}\ \bibnamefont
  {{Geraci}}}, \bibinfo {author} {\bibfnamefont {S.~B.}\ \bibnamefont
  {{Papp}}}, \ and\ \bibinfo {author} {\bibfnamefont {J.}~\bibnamefont
  {{Kitching}}},\ }\href {\doibase 10.1103/PhysRevLett.105.101101} {\bibfield
  {journal} {\bibinfo  {journal} {Phys. Rev. Lett.}\ }\textbf {\bibinfo
  {volume} {105}},\ \bibinfo {eid} {101101} (\bibinfo {year} {2010})},\ \Eprint
  {http://arxiv.org/abs/1006.0261} {arXiv:1006.0261 [hep-ph]} \BibitemShut
  {NoStop}%
\bibitem [{\citenamefont {Moore}\ \emph {et~al.}(2014)\citenamefont {Moore},
  \citenamefont {Rider},\ and\ \citenamefont {Gratta}}]{Moore14}%
  \BibitemOpen
  \bibfield  {author} {\bibinfo {author} {\bibfnamefont {D.~C.}\ \bibnamefont
  {Moore}}, \bibinfo {author} {\bibfnamefont {A.~D.}\ \bibnamefont {Rider}}, \
  and\ \bibinfo {author} {\bibfnamefont {G.}~\bibnamefont {Gratta}},\ }\href
  {\doibase 10.1103/PhysRevLett.113.251801} {\bibfield  {journal} {\bibinfo
  {journal} {Phys. Rev. Lett.}\ }\textbf {\bibinfo {volume} {113}},\ \bibinfo
  {pages} {251801} (\bibinfo {year} {2014})},\ \Eprint
  {http://arxiv.org/abs/1408.4396} {arXiv:1408.4396 [hep-ex]} \BibitemShut
  {NoStop}%
%%CITATION = ARXIV:1408.4396;%%
\bibitem [{\citenamefont {{Gieseler}}\ \emph {et~al.}(2013)\citenamefont
  {{Gieseler}}, \citenamefont {{Novotny}},\ and\ \citenamefont
  {{Quidant}}}]{Gieseler13}%
  \BibitemOpen
  \bibfield  {author} {\bibinfo {author} {\bibfnamefont {J.}~\bibnamefont
  {{Gieseler}}}, \bibinfo {author} {\bibfnamefont {L.}~\bibnamefont
  {{Novotny}}}, \ and\ \bibinfo {author} {\bibfnamefont {R.}~\bibnamefont
  {{Quidant}}},\ }\href {\doibase 10.1038/nphys2798} {\bibfield  {journal}
  {\bibinfo  {journal} {Nat. Phys.}\ }\textbf {\bibinfo {volume} {9}},\
  \bibinfo {pages} {806} (\bibinfo {year} {2013})},\ \Eprint
  {http://arxiv.org/abs/1307.4684} {arXiv:1307.4684 [cond-mat.mes-hall]}
  \BibitemShut {NoStop}%
\bibitem [{\citenamefont {{Ranjit}}\ \emph {et~al.}(2016)\citenamefont
  {{Ranjit}}, \citenamefont {{Cunningham}}, \citenamefont {{Casey}},\ and\
  \citenamefont {{Geraci}}}]{Ranjit16}%
  \BibitemOpen
  \bibfield  {author} {\bibinfo {author} {\bibfnamefont {G.}~\bibnamefont
  {{Ranjit}}}, \bibinfo {author} {\bibfnamefont {M.}~\bibnamefont
  {{Cunningham}}}, \bibinfo {author} {\bibfnamefont {K.}~\bibnamefont
  {{Casey}}}, \ and\ \bibinfo {author} {\bibfnamefont {A.~A.}\ \bibnamefont
  {{Geraci}}},\ }\href@noop {} {\  (\bibinfo {year} {2016})},\ \Eprint
  {http://arxiv.org/abs/1603.02122} {arXiv:1603.02122 [physics.optics]}
  \BibitemShut {NoStop}%
\bibitem [{Note1()}]{Note1}%
  \BibitemOpen
  \bibinfo {note} {Bangs Laboratories, Inc.,
  http://www.bangslabs.com}\BibitemShut {NoStop}%
\bibitem [{\citenamefont {{Ashkin}}\ and\ \citenamefont
  {{Dziedzic}}(1971)}]{Ashkin71}%
  \BibitemOpen
  \bibfield  {author} {\bibinfo {author} {\bibfnamefont {A.}~\bibnamefont
  {{Ashkin}}}\ and\ \bibinfo {author} {\bibfnamefont {J.~M.}\ \bibnamefont
  {{Dziedzic}}},\ }\href {\doibase 10.1063/1.1653919} {\bibfield  {journal}
  {\bibinfo  {journal} {Appl. Phys. Lett.}\ }\textbf {\bibinfo {volume} {19}},\
  \bibinfo {pages} {283} (\bibinfo {year} {1971})}\BibitemShut {NoStop}%
\bibitem [{\citenamefont {{Ashkin}}\ \emph {et~al.}(1986)\citenamefont
  {{Ashkin}}, \citenamefont {{Dziedzic}}, \citenamefont {{Bjorkholm}},\ and\
  \citenamefont {{Chu}}}]{ashkin}%
  \BibitemOpen
  \bibfield  {author} {\bibinfo {author} {\bibfnamefont {A.}~\bibnamefont
  {{Ashkin}}}, \bibinfo {author} {\bibfnamefont {J.~M.}\ \bibnamefont
  {{Dziedzic}}}, \bibinfo {author} {\bibfnamefont {J.~E.}\ \bibnamefont
  {{Bjorkholm}}}, \ and\ \bibinfo {author} {\bibfnamefont {S.}~\bibnamefont
  {{Chu}}},\ }\href {\doibase 10.1364/OL.11.000288} {\bibfield  {journal}
  {\bibinfo  {journal} {Opt. Lett.}\ }\textbf {\bibinfo {volume} {11}},\
  \bibinfo {pages} {288} (\bibinfo {year} {1986})}\BibitemShut {NoStop}%
\bibitem [{Note2()}]{Note2}%
  \BibitemOpen
  \bibinfo {note} {Newport, product number: NPXYZ100SGV6,
  http://www.newport.com}\BibitemShut {NoStop}%
\bibitem [{\citenamefont {Jackson}(1999)}]{jackson99}%
  \BibitemOpen
  \bibfield  {author} {\bibinfo {author} {\bibfnamefont {J.~D.}\ \bibnamefont
  {Jackson}},\ }\href {http://cdsweb.cern.ch/record/490457} {\emph {\bibinfo
  {title} {Classical electrodynamics}}},\ \bibinfo {edition} {3rd}\ ed.\
  (\bibinfo  {publisher} {Wiley},\ \bibinfo {address} {New York, {NY}},\
  \bibinfo {year} {1999})\BibitemShut {NoStop}%
\bibitem [{\citenamefont {Wilks}(1938)}]{wilks38}%
  \BibitemOpen
  \bibfield  {author} {\bibinfo {author} {\bibfnamefont {S.~S.}\ \bibnamefont
  {Wilks}},\ }\href {\doibase 10.1214/aoms/1177732360} {\bibfield  {journal}
  {\bibinfo  {journal} {Ann. Math. Statist.}\ }\textbf {\bibinfo {volume}
  {9}},\ \bibinfo {pages} {60} (\bibinfo {year} {1938})}\BibitemShut {NoStop}%
\bibitem [{\citenamefont {Cowan}(1998)}]{Cowan98}%
  \BibitemOpen
  \bibfield  {author} {\bibinfo {author} {\bibfnamefont {G.}~\bibnamefont
  {Cowan}},\ }\href {https://books.google.com/books?id=ff8ZyW0nlJAC} {\emph
  {\bibinfo {title} {Statistical Data Analysis}}},\ Oxford science
  publications\ (\bibinfo  {publisher} {Clarendon Press},\ \bibinfo {year}
  {1998})\BibitemShut {NoStop}%
\bibitem [{\citenamefont {Adelberger}\ \emph {et~al.}(2007)\citenamefont
  {Adelberger}, \citenamefont {Heckel}, \citenamefont {Hoedl}, \citenamefont
  {Hoyle}, \citenamefont {Kapner},\ and\ \citenamefont
  {Upadhye}}]{Adelberger06}%
  \BibitemOpen
  \bibfield  {author} {\bibinfo {author} {\bibfnamefont {E.~G.}\ \bibnamefont
  {Adelberger}}, \bibinfo {author} {\bibfnamefont {B.~R.}\ \bibnamefont
  {Heckel}}, \bibinfo {author} {\bibfnamefont {S.~A.}\ \bibnamefont {Hoedl}},
  \bibinfo {author} {\bibfnamefont {C.~D.}\ \bibnamefont {Hoyle}}, \bibinfo
  {author} {\bibfnamefont {D.~J.}\ \bibnamefont {Kapner}}, \ and\ \bibinfo
  {author} {\bibfnamefont {A.}~\bibnamefont {Upadhye}},\ }\href {\doibase
  10.1103/PhysRevLett.98.131104} {\bibfield  {journal} {\bibinfo  {journal}
  {Phys. Rev. Lett.}\ }\textbf {\bibinfo {volume} {98}},\ \bibinfo {pages}
  {131104} (\bibinfo {year} {2007})},\ \Eprint
  {http://arxiv.org/abs/hep-ph/0611223} {arXiv:hep-ph/0611223} \BibitemShut
  {NoStop}%
%%CITATION = HEP-PH/0611223;%%
\bibitem [{\citenamefont {{Bishop}}\ \emph {et~al.}(2004)\citenamefont
  {{Bishop}}, \citenamefont {{Nieminen}}, \citenamefont {{Heckenberg}},\ and\
  \citenamefont {{Rubinsztein-Dunlop}}}]{bishop2004}%
  \BibitemOpen
  \bibfield  {author} {\bibinfo {author} {\bibfnamefont {A.~I.}\ \bibnamefont
  {{Bishop}}}, \bibinfo {author} {\bibfnamefont {T.~A.}\ \bibnamefont
  {{Nieminen}}}, \bibinfo {author} {\bibfnamefont {N.~R.}\ \bibnamefont
  {{Heckenberg}}}, \ and\ \bibinfo {author} {\bibfnamefont {H.}~\bibnamefont
  {{Rubinsztein-Dunlop}}},\ }\href {\doibase 10.1103/PhysRevLett.92.198104}
  {\bibfield  {journal} {\bibinfo  {journal} {Phys. Rev. Lett.}\ }\textbf
  {\bibinfo {volume} {92}},\ \bibinfo {eid} {198104} (\bibinfo {year}
  {2004})},\ \Eprint {http://arxiv.org/abs/physics/0402021}
  {arXiv:physics/0402021} \BibitemShut {NoStop}%
\bibitem [{\citenamefont {Simpson}\ \emph {et~al.}(2002)\citenamefont
  {Simpson}, \citenamefont {Wilson}, \citenamefont {Gericke},\ and\
  \citenamefont {Zare}}]{simpson2002}%
  \BibitemOpen
  \bibfield  {author} {\bibinfo {author} {\bibfnamefont {G.~J.}\ \bibnamefont
  {Simpson}}, \bibinfo {author} {\bibfnamefont {C.~F.}\ \bibnamefont {Wilson}},
  \bibinfo {author} {\bibfnamefont {K.-H.}\ \bibnamefont {Gericke}}, \ and\
  \bibinfo {author} {\bibfnamefont {R.~N.}\ \bibnamefont {Zare}},\ }\href
  {\doibase 10.1002/1439-7641(20020517)3:5<416::AID-CPHC416>3.0.CO;2-K}
  {\bibfield  {journal} {\bibinfo  {journal} {ChemPhysChem}\ }\textbf {\bibinfo
  {volume} {3}},\ \bibinfo {pages} {416} (\bibinfo {year} {2002})}\BibitemShut
  {NoStop}%
\bibitem [{\citenamefont {{Millen}}\ \emph {et~al.}(2014)\citenamefont
  {{Millen}}, \citenamefont {{Deesuwan}}, \citenamefont {{Barker}},\ and\
  \citenamefont {{Anders}}}]{Millen14}%
  \BibitemOpen
  \bibfield  {author} {\bibinfo {author} {\bibfnamefont {J.}~\bibnamefont
  {{Millen}}}, \bibinfo {author} {\bibfnamefont {T.}~\bibnamefont
  {{Deesuwan}}}, \bibinfo {author} {\bibfnamefont {P.}~\bibnamefont
  {{Barker}}}, \ and\ \bibinfo {author} {\bibfnamefont {J.}~\bibnamefont
  {{Anders}}},\ }\href {\doibase 10.1038/nnano.2014.82} {\bibfield  {journal}
  {\bibinfo  {journal} {Nat. Nano.}\ }\textbf {\bibinfo {volume} {9}},\
  \bibinfo {pages} {425} (\bibinfo {year} {2014})},\ \Eprint
  {http://arxiv.org/abs/1309.3990} {arXiv:1309.3990 [cond-mat.soft]}
  \BibitemShut {NoStop}%
\end{thebibliography}%

\end{document}